\documentclass[epsfig]{aipproc}
\layoutstyle{8x11double}

\begin{document}

\def\gr{\hbox{ \raisebox{-1.0mm}{$\stackrel{>}{\sim}$} }}
\def\kr{\hbox{ \raisebox{-1.0mm}{$\stackrel{<}{\sim}$} }}

\title{Gamma-Ray Bursts and Cosmic Radiation Backgrounds}


\author{D. H. Hartmann}{
  address={Department of Physics and Astronomy, Clemson University, 
  Clemson, SC 29634-0978, USA}}

\author{T. M. Kneiske and K. Mannheim}{
  address={Universit\"at W\"urzburg, Am Hubland, 97057 W\"urzburg, Germany}}

\author{K. Watanabe}{
  address={SSAI/LHEA, NASA/GSFC, Code 664, Greenbelt, MD 20771, USA}
  }

\copyrightyear  {2002}

\begin{abstract}
If gamma-ray bursts trace the cosmic star formation rate to large redshifts,
their prompt and delayed emissions provide new tools for early universe cosmology.
In addition to probing the intervening matter via absorption lines in the optical
band, GRB continua also contribute to the evolving cosmic radiation background.
We discuss the contribution of GRBs to the high-energy background, and the effect
pair creation off low-energy background photons has on their observable TeV spectra. 
\end{abstract}

\date{\today}

\maketitle

                   \section{Introduction}
Cosmic gamma-ray bursts (GRBs) have redshifts comparable to or perhaps even
larger than those of quasars. Indeed, they are the most energetic explosions
in the universe, with energies (uncorrected for beaming) of order M$_\odot$c$^2$.
Their host galaxies are often sub-L$_*$, but actively forming stars at rates
typical for galaxies in the early universe. The current paradigm associates
GRBs with the formation of black holes in massive, rapidly rotating stars,
or with the merger of compact star binaries. GRBs thus trace directly, or
perhaps with a short delay, the cosmic star formation history, and may be the
most easily detectable signposts of the first generation of stars (e.g., [1]
and references therein). The redshifted gamma-ray flux from GRBs contributes 
to the evolving radiation background of the universe, as discussed in the next
section, and at the same time serves as a probe of the cosmic radiation field
through electron-positron pair creation absorption of their highest energy
photons [2][3]. 

The cosmic microwave background (CMB) provides abundant soft photons for 
pair production of very high energy photons (in the TeV - PeV regime), but 
at lower energies (GeV-TeV regime) the target photons are optical and IR
photons produced by stars and reprocessed by surrounding dust. Cosmic
chemical evolution is intimately linked to the cosmic star formation history,
and the present day extragalactic background light (EBL) provides a record
of that history. Gamma-ray sources, such as GRBs and blazars, probe the 
evolution of this photon field through absorption effects at high energies
(e.g., [2][3]). There are only three nearby active galaxies for which this
absorption effect has been observed, Mrk 421, Mrk 501 (both at z = 0.03), and
BL Lac (at z = 0.044).
TeV emission from GRBs has only been reported for GRB970417a [4].
GRB power spectra ($\nu$f$_\nu$) typically peak at photon energies of a few 
hundred keV, but their power-law high energy emision may extend well into the
GeV or even TeV regime. EGRET aboard the Compton Observatory has established
that emission above 100 MeV is common, and in the case of  
GRB940217 a maximum photon energy of E $\sim$ 20 GeV was determined [5].

Theoretical models (e.g., [6]) certainly suggest that GeV-TeV emission
should be expected for a significant fraction of all bursts. The next generation
GLAST experiment is expected to observe a large number of GRBs with spectral
coverage up to 300 GeV. Ongoing improvements of ground-based experiments
(VERITAS, HESS, HEGRA, MILAGRO, MAGIC, ...) lead to reduced sensitivities and
thresholds, thus overlapping with space-based experiments. It will thus be 
possible to explore GRB spectra from the X-ray regime to the TeV regime, and 
the effects of propagation effects such as the above mentioned electron-positron
pair creation must be taken into account. 

To correct for $\gamma\gamma$
absorption, it is necessary to determine the cosmic evolution of the target 
photon distribution function, which we refer to as the metagalactic radiation
field (MRF). In the third section of this paper we briefly describe
our simulations of the evolving low-energy MRF, and 
demonstrate the extinction effect in the high energy part of GRB spectra. The 
gamma-ray horizon of the universe can perhaps be probed with GRBs, which would
provide another powerful tool for the study of stellar evolution on the
cosmic scale. GRB detections point to the onset of star formation in the
universe, and their high energy spectra probe the production of light 
throughout the cosmic ages.

     \section{The Gamma-Ray Background}

The unresolved cosmic gamma-ray background (CGB) from 10 KeV to 100 GeV is 
predominantly due to the superposition of three source populations (e.g., [7] 
and references therein): Seyfert galaxies, which dominate below $\sim$ 100
keV; blazars, which dominate above a few MeV, and Type Ia supernovae, which
fill the gap between the contributing active galaxies. The flux in the MeV
regime, detected with COMPTEL and SMM, can be accounted for with nuclear 
line emission from the decay chain $^{56}$Ni $\rightarrow$ $^{56}$Co $\rightarrow$
$^{56}$Fe, escaping from expanding supernova ejecta (e.g. [8]). Despite their
lower rate SNIa contribute the bulk of the CGB in the MeV regime due to their
higher yields, and low mass envelopes. Core collapse supernovae (SNII) do
contribute about 1$\%$ of the flux, as do radio galaxies such as Cen A ([9]).
Figure 1 shows the observations and the contributions from various sources.
The fact that SNIa so closely match the observed spectrum of the CGB places
significant constraints on the cosmic star formation rate SFR(z). Most of 
observed flux is due to the integrated emission from supernovae that
exploded at redshifts less than $\sim$ 1, but a SFR that continues to rise
as rapidly as observed between the present epoch and z$\sim$1 would overproduce
the CGB. The MeV data thus provide an independent constraint on the high-z
star formation history of the universe.

\begin{figure}[t!]
\resizebox{18pc}{!}{\includegraphics{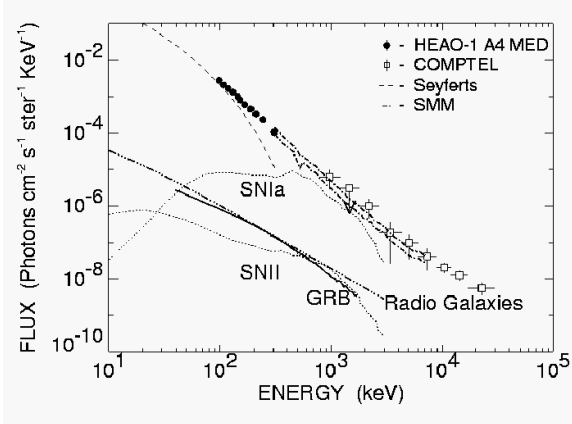}}
\caption{The observed gamma-ray background and estimated contributions from
supernovae, radio galaxies, and gamma-ray bursts [7]$-$[9].} 
\end{figure}

Throughout the observable universe GRBs occur at significantly lower rates than
supernovae, but their high $\gamma$-ray luminosity suggests a possible contribution
to the CGB. Watanabe and Hartmann [9] used the BATSE data from the 4B catalog
to estimate the contribution of observed GRBs (corrected for Earth blocking)
to the CGB. As Figure 1 shows, GRBs compete with radio galaxies and SNII, but
do not contribute a large portion of the observed background. Still, their
role could be significant by reducing some of the deficit at 200-400 keV
between the observed spectrum and the predicted flux from Seyfert galaxies.

      \section{The Low-Energy Background}

Numerical simulations of hierarchical
structure formation in a globally homogeneous Universe are tractable, but
connecting the evolving gravitational structures to observable fluxes
of electromagnetic radiation involves uncertain empirical descriptions of star
formation, supernova feedback, and the dust-gas-star interplay. The necessary input
comes from extensive observational campaigns, such as deep galaxy surveys, which
measure the number of galaxies, morphological types, colors, fluxes, and distances
in presumably representative solid angles out to large redshifts. The wealth of
information derived from these observations can significantly complicate efforts 
to link theories of galaxy evolution and large scale structure formation. It is
helpful to single out key quantities for which predictions can be compared with
observations. One such quantity is the cosmic star formation rate (SFR) and its
associated metagalactic radiation field (MRF).
The MRF at $z=0$ is commonly referred to as Extragalactic Background Light (EBL).
The contribution of galaxies to the
MRF is most significant between the far-infrared and the ultraviolet, while at longer
wavelengths the 2.7~K microwave background (CMB) radiation from the big bang dominates.
 
In principle, the evolution of the MRF should be predictable from structure formation
models [10] so that the observed MRF could be used
to infer the role of AGNs, low surface brightness objects, and the decays
of relic particles, or to single out global cosmological parameters.
However, these models still rely on many uncertain parameters, and
we are far from the ultimate goal of a first principles theory of the MRF.
Here we compute the MRF directly from the global SFR inferred from tracers
of cosmic chemical evolution, such as various Lyman $\alpha$ absorber systems,
or from deep galaxy surveys.
The spectral energy distribution (SED) for the globally averaged stellar population
residing in galaxies can be estimated with population synthesis models [11]
available for various input parameters, of which the
initial mass function (IMF) and metallicity ($Z$) are the most important ones.
Reprocessing by gas and dust is taken into account explicitly via some
model of the evolution of the dust and gas content in galaxies, in combination
with assumed dust properties derived from local observations in the Milky Way.
The details of our modeling of the MRF are presented in [12].

Observational attempts to determine or constrain the present-day background face
severe problems due to emissions from the Galaxy, which can introduce large
systematic errors [13].
Nevertheless, studies with COBE
have resulted in highly significant
detections of a residual diffuse IR background,
providing an upper bound on the MRF in the IR regime. Similarly, the cumulative
flux from galaxies detected in deep HST or ISO exposures provide useful lower 
limits to the present-day MRF (e.g., [13]). 

The method for calculating the MRF from a given SFR relies on an accurate knowledge of
evolving stellar spectra and the reprocessing of star light in various dusty environments.
Luminosity evolution of stellar populations is sensitive to the IMF,
evolution of the mean cosmic metallicity, and the amount of interstellar extinction.
Starting point of any model is the spectral energy density (SED) produced by a population
of stars resulting from an instantaneous burst of star formation (commonly
normalized to the mass of stars formed). Because star formation is an ongoing process
with relatively short time scales of 10$^{5-7}$ yrs, the starburst spectra can be directly
convolved with the global SFR, $\dot{\rho_\ast}(z)$, to derive the evolution of the global
luminosity density due to cosmic star formation. The SEDs are
constructed from realistic stellar evolution tracks combined with detailed atmospheric
models (e.g., [11]). The temporal evolution of the specific
luminosity, $L_\nu(t)$ (in erg~s$^{-1}$Hz$^{-1}$ per unit mass of stars formed)
is then determined by the choices of IMF and the initial stellar metallicity.

From the population synthesis starburst models we obtain the comoving emissivity 
(luminosity density) at cosmic epoch $t$ from the convolution
\begin{equation}
\mathcal{E}_{\nu}(t) = \int_{t_m}^{t}
L_{\nu}(t-t')\dot{\tilde\rho}_{\ast}(t') dt' ~ (\mathrm{erg s^{-1} Hz^{-1} Mpc^{-1}}) ~
\label{eq:emist}
\end{equation}
where $\dot{\tilde\rho}_\ast(t)=\dot{\rho}_\ast(z)$ is the star formation rate per comoving
unit volume. Rewriting Eq.~(\ref{eq:emist}) in terms of redshift, $z=z(t)$, yields
\begin{equation}
\mathcal{E}_{\nu}(z) = \int_z^{z_{m}}
L_{\nu}(t(z)-t(z'))\dot{\rho}_{\ast}(z') \left | \frac{dt'}{dz'} \right |
dz' ~,
\label{eq:emislambda}
\end{equation}
where we assumed that star formation began at some finite epoch
$z_m=z(t_m)$. For given evolution of the emissivity a second integration over
redshift yields the energy density, or, after multiplication with $c/4\pi$, the
comoving power spectrum of the MRF
\begin{equation} P_\nu(z) = \nu
I_{\nu}(z) = \nu \frac{c}{4\pi} \int_z^{z_m}  \mathcal{E}_{\nu'}(z')
\left | \frac{dt'}{dz'} \right |  dz' ~ , \label{eq:hinter} \end{equation}
with $\nu'=\nu(1+z')/(1+z)$.
Cosmological parameters enter through $dt/dz$,
given by 
\begin{equation}
\left | \frac{dt}{dz} \right | =\frac{1}{H_0(1+z)E(z)}
\end{equation}
with an equation of state
\begin{equation}
E(z)^2=
\Omega_{\rm r}
(1+z)^4+\Omega_{\rm m}(1+z)^3+\Omega_R(1+z)^2+\Omega_\Lambda ~.
\end{equation}
The term proportional to $\Omega_{\rm r}$ takes into account the contribution from relativistic
components such as the CMB and star light, although the latter would also require a new function 
describing the production of light as a function of time. The density parameter of this component 
is defined as $\Omega_{\rm r} =
u_{\rm r}/\rho_{\rm crit}c^2$, where
$u_{\rm r}$ refers to the relativistic energy density and $\rho_{\rm crit}$ is the
critical density of the universe; $\rho_{\rm crit}$ = 3H$_0^2$/8$\pi$G = 10.54 h$^2$ keV/cm$^3$.

The average metallicty of gas in galaxies slowly increases with cosmic time, but
the present-day value is not known precisely (e.g., [14]).
We thus adopt an average extinction curve
\begin{equation}
A_\lambda=0.68\cdot E(B-V)\cdot R\cdot (\lambda^{-1}-0.35)
\end{equation}
with $R=3.2$ and
where $A_\lambda$ with $\lambda$ [$\mu$m] determines the absorption coefficient 
according to $g(\lambda)=10^{-0.4\cdot A_{\lambda}}$.
Reemission by dust is calculated as the sum of three modified Planck spectra
\begin{equation}
L_{\lambda}^d(L_{bol}) = \sum_{i=1}^{3} c_i(L_{bol})\cdot Q_\lambda \cdot B_\lambda(T_i)
\label{eq:planck} \end{equation}
where
$Q_\lambda \propto \lambda^{-1}$.
Two temperatures characterize warm and cold dust in galaxies, and one temperature is
included to emulate a PAH component, which is also assumed to emit like a Blackbody.
Dust in the ISM of the Milky Way is known to coexist at several different
temperatures, determined by the distances from various heat sources. Hot dust 
has temperatures ranging 
from 50~K to 150~K-200~K 
when the dust is in equilibrated
within HII regions, or near massive stars or compact accreting sources. Radiation 
from this dust component predominantly
emerges in the mid-infrared and reprocesses only a small fraction of the stellar
luminosity. Warm dust with temperatures between 25~K and 50~K corresponds to
regions heated by the mean interstellar radiation field. Dust inside molecular clouds
is shielded against high-energy radiation, and thus appears at low
temperatures between 10~K and 25~K. Very cold dust at temperatures of 10~K or less 
can be present in the densest parts of molecular clouds or in outer regions of
the galaxy where the flux of the interstellar radiation field has dropped to
the value of the MRF.

The cosmic star formation rate density SFR(z) has been determined with
different methods and for large set of input data. Many of these studies suggest
that the original Madau curve [15] should be considered a lower limit,
and that realistic rates could be
larger by a factor 2$-$3 at all redshifts. A review of published SFR(z) functions  
shows that we
do not yet understand systematic effects well enough to obtain a reliable
estimate for SFR(z). This is especially true at redshifts beyond unity.
This uncertainty enters in the final step of computing the MRF, the integration of 
the emissivity  over cosmic time using Eq.~(\ref{eq:hinter}).
The evolution of the resulting MRF spectrum is shown
in Fig. 2 for several redshifts.

\begin{figure}[t!]
\resizebox{18pc}{!}{\includegraphics{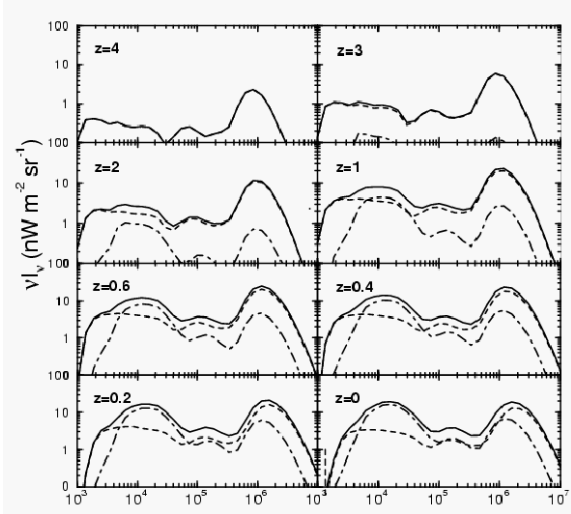}}
\caption{
The evolving spectrum of the extragalactic background light.
{\em Dashed} lines show the contribution of massive stars and 
{\em dot-dashed} lines the contribution of low mass stars} 
\end{figure}

      \section{Absorbed TeV spectra}

Gamma-ray absorption due to $\gamma\gamma$-pair creation on cosmological scales
depends on the line-of-sight integral of the evolving density of low-energy photons
in the Universe, i.e. on the history of the diffuse, isotropic radiation field.
Above we briefly discussed our semi-empirical MRF model, which is 
based on stellar light produced and reprocessed in evolving galaxies and 
calibrated with the EBL.
The optical depth of the universe is given by
\begin{equation}
\tau_{\gamma\gamma}(E,z) = \int_0^z\ dz (\partial_zl)\ n(\epsilon,z)\ 
\left<\sigma(E,z)\right>
\end{equation} 
where E is the energy of the observed photon, n($\epsilon$,z) represents the evolving MRF, 
and the angle averaged cross section $\left<\sigma\right>$ is of order
of the Thomson cross section, $\sigma_T$. Using a power law spectrum as template
(with an intrinsic absorption above 1 TeV) we show in Figure 3 how the 
line of sight optical depth to $\gamma\gamma$-pair creation affects the spectra.
It is apparent that most GRB spectra, if they intrinsically extend into the TeV
regime, are severely affected as soon as their redshifts exceed z $\sim$ 0.1. If 
the GRB distribution traces the cosmic star formation history, we expect only a 
a small fraction of all bursts to be close enough for detectable TeV emission. Ongoing 
efforts to observe TeV emission from GRBs have so far only turned up one possible 
detection (GRB 970417a), which suggests either that GRBs do not commonly radiate 
in this regime, or that they do but are extinct by the opacity along the line of
sight. If the latter interpretation is correct, TeV detections of GRBs (for example 
with ground based muon detectors; [16]) will be rare [2]
but valuable probes of GRB physics and MRF evolution. TeV GRBs could significantly
enhance the insights gathered from the limited set of TeV blazars (e.g., [17])

\begin{figure}[t!]
\resizebox{18pc}{!}{\includegraphics{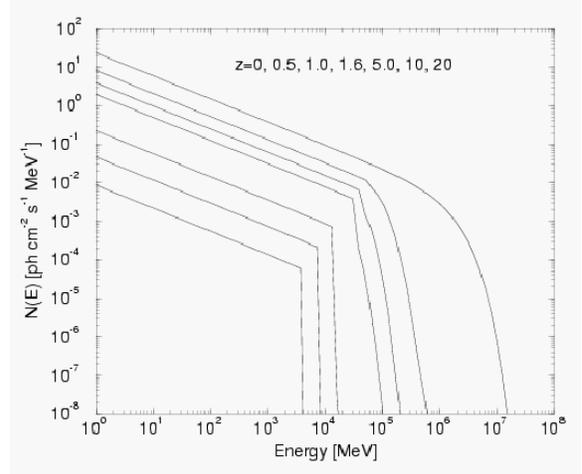}}
\caption{Using a power-law spectrum as a template (with
an intrinsic cut-off above E $\sim$ 1 TeV), the figure shows
the effect of pair creation absorption for various redshifts.}
\end{figure}




\begin{thebibliography}{12}

\bibitem{[1]} Hartmann, D. H., MacFadyen, A. I., and Woosley, S. E.
              in \it Gamma-Ray Bursts\rm, eds. 
              R. M. Kippen, R. S. Mallozzi, and G. J. Fishman, 
              AIP 526, p. 653.
\bibitem{[2]} Mannheim, K., Hartmann, D. H., and Funk, B. 1996, ApJ 467, 532.
\bibitem{[3]} Salamon, M. H., and Stecker, F. W. 1998, ApJ 493, 547
\bibitem{[4]} Atkins, R., et al. 2000, ApJ 533, L119
\bibitem{[5]} Hurley, K., et al. 1994, {\it Nature} 372, 652
\bibitem{[6]} Zhang, B., and Meszaros, P. 2001, ApJ 559, 110
\bibitem{[7]} Watanabe, K., et al. 1999, ApJ 516, 285
\bibitem{[8]} The, L.-S. et al. 1993, ApJ 403, 32
\bibitem{[9]} Watanabe, K. and Hartmann, D. H.
              in \it Gamma 2001\rm, eds.
              S. Ritz, N. Gehrels, and C. R. Shrader
              AIP 587, p. 442.
\bibitem{[10]} Somerville, R. \& Primack, J.R. 1999 MNRAS 310, 1087
\bibitem{[11]} Bruzual, A. G., $\&$ Charlot, S. 1993, ApJ 405, 538
\bibitem{[12]} Kneiske, T. M., Mannheim, K., and Hartmann, D. H. 2002, A$\&$A, submitted
\bibitem{[13]} Bernstein, R.A., Freedman, W. , Madore, B.F. 2001, astro-ph/0112153
\bibitem{[14]} Pei, Y.C., Fall, S.M., \& Hauser, M.G. 1999, ApJ 522, 604.
\bibitem{[15]} Madau, P. 1997, ApJ 475, 429. 
\bibitem{[16]} Gupta, N. $\&$ Bhattacharjec, P. 2001, astro-ph/0108311
\bibitem{[17]} Aharonian, F. A., 2001, astro-ph/0112314

\end{thebibliography}
\end{document}